\documentstyle[12pt,epsfig]{article}
 
\begin{document}
\pagestyle{empty}

\noindent
\hfill
\begin{minipage}[t]{3in}
\begin{flushright}
UCB--PTH--96/37 \\
LBNL--39399\\
September 1996
\end{flushright}
\end{minipage}
\begin{center}

\vskip 1cm

{\large \bf Axial-vector meson mixing in orthocharmonium decays}

\vskip 1cm

Mahiko Suzuki\\

{\em Department of Physics and Lawrence Berkeley National Laboratory\\
University of California, Berkeley, California 94720
}

\end{center}



\vskip 1cm

\begin{abstract}
In the light of the recent measurement by the BES Collaboration, 
the two-body decays of $J/\psi$ and $\psi'$ 
into an axial-vector meson and a pseudoscalar meson are analyzed 
in the framework of the $K_A-K_B$ mixing including substantial 
SU(3) and G parity violations due to one-photon annihilation.
A somewhat puzzling pattern of the $K_1^+K^-$ decay channel 
can be understood with no tight constraint on the mixing angle. The ratio of 
$K_1^0(1400)\overline{K}^0$ to $K_1^0(1270)\overline{K}^0$ 
will be the cleanest source of information to determine the mixing angle 
from the $1^+0^-$ decays in the presence of one-photon annihilation.

\end{abstract}
\newpage
\pagestyle{plain}

\setcounter{footnote}{0}

\section{Introduction}
   Mixing between the strange meson states of two axial-vector octets
was established through their decay modes, mass splitting, and production
in the $\tau$ decay \cite{old,suzuki,new}.  In the zeroth order
approximation, the mixing is maximal within large uncertainties  
according to the current experimental information.
   The BES Collaboration recently reported on the $1^+0^-$ decays of the
orthocharmonia \cite{BES}.  In their measurement, the branching fraction 
Br$(\psi'\rightarrow K_1^+(1270)K^-)$ is far dominant over 
Br$(\psi'\rightarrow K_1^+(1400)K^-)$, while it is in the other way
in the $J/\psi$ decay. If one ignores one-photon annihilation 
and assumes the maximal mixing, the $K_1(1270)\overline{K}$ 
and $K_1(1400)\overline{K}$ branching fractions would be 
equal to each other both in $J/\psi$ and $\psi'$ decays. 
We study here implications of the BES data on the $K_A-K_B$ mixing. 
The purpose of this paper is twofold: First to show that the BES 
measurement of the $K_1^+K^- +cc$ \cite{BES} is perfectly consistent with 
the $K_A-K_B$ mixing when one-photon annihilation is properly added to 
three-gluon annihilation, and second to show that the cleanest determination 
of the mixing angle from the $1^+0^-$ decay of orthcharmonia is to measure
the neutral modes of $K_1\overline{K}$. 

Our analysis is based on flavor SU(3) symmetry with the following 
standard assumptions.

(1) The decay occurs through three-gluon annihilation and also through
one-photon annihilation. While the former is dominant over the latter, 
their relative magnitude is left as a parameter.

(2) The strange meson components of the axial-vector octets 
mix with each other. The mixing angle is roughly $45^{\circ}$, 
but with large uncertainties.

(3) Flavor SU(3) invariance is valid for strong interactions apart from
the meson mixing. The electromagnetic current of the light quarks is 
an octet with negative charge conjugation.

\section{SU(3) parametrization}

   Two axial-vector meson octets have been known.  They form approximate 
nonets like the vector mesons though one state is still missing:
\begin{equation}
    a_1(1260),\; K_A,\; \overline{K}_A, \; f_1(1285),\; f'_1(1420)\;\;  \cdots 
                   1^+_A({\bf 8})-1^+_A({\bf 1}),
\end{equation}
\begin{equation}
    b_1(1235),\; K_B, \;\overline{K}_B,\; h_1(1170),\; h'_1(?) \;\;\;\; \cdots
                    1^+_B({\bf 8})-1^+_B({\bf 1}),
\end{equation}
where the numbers in boldface are SU(3) representations.  $K_A$ and $K_B$ 
mix through SU(3) breaking to form mass eigenstates $K_1(1270)$
and $K_1(1400)$ as
\begin{eqnarray}
    K_1(1400)  &=& K_A \cos\theta - K_B \sin\theta, \nonumber\\
    K_1(1270)  &=& K_A \sin\theta + K_B \cos\theta.  \label{mixing}
\end{eqnarray}
Similarly for $\overline{K_1}$ with $\theta\rightarrow -\theta$.
The mixing coefficients in Eq.(\ref{mixing}) are real provided that the 
dispersive part dominates over the absorptive part in the mass matrix.
It should be noted that the signs of $\sin\theta$ and $\cos\theta$ 
can be absorbed into the phases of particle states or fields.  Hereafter
we shall choose the phase convention of states 
such that
\begin{equation}
              0^{\circ} \leq \theta \leq 90^{\circ}.
\end{equation}

   Since $1^+_A({\bf 8})$, $1^+_A({\bf 1})$ or $1^+_B({\bf 1})$ plus 
a pseudoscalar octet $0^-({\bf 8})$ cannot form an SU(3) singlet of 
negative charge conjugation, the three-gluon annihilation allows only   
\begin{equation}
   J/\psi, \;\;\; \psi' \rightarrow  1^+_B({\bf 8})\;0^-({\bf 8}). \label{M0} 
\end{equation}
In contrast, the one-photon annihilation allows most of $1^+0^-$:
\begin{equation}
     J/\psi,\;\;\psi'  \rightarrow  1^+_A({\bf 8})\;0^-({\bf 8}),\;\;
                                    1^+_B({\bf 8})\;0^-({\bf 8}),\;\;
                                    1^+_B({\bf 1})\;0^-({\bf 8}).  \label{Mv}
\end{equation}
Let us denote the decay amplitude of Eq.(\ref{M0}) by $M_0$, and the 
amplitudes for the first and second processes of Eq.(\ref{Mv}) by $M_A$ and 
$M_B$, respectively. Then we can describe the decay branching fractions of 
$J/\psi$ or $\psi'$ into $1^+({\bf 8})\,0^-({\bf 8})$ by three parameters,
\begin{equation}
   \theta, \;\;\;\; \xi\equiv M_A/M_0, 
           \;\;\;\; \eta\equiv M_B/M_0.
\end{equation}
We distinguish the parameters $\xi$ and $\eta$ for $\psi'$ from 
those for $J/\psi$ by attaching them primes
($\xi\rightarrow\xi'$ and $\eta\rightarrow\eta'$).

  The standard SU(3) analysis gives the 
decay amplitudes as shown in Table 1. For the channels involving $h_1$ 
or $h'_1$, the $1_B^+({\bf 1})\,0^-({\bf 8})$ coupling has been related to the
$1_B^+({\bf 8})\,0^-({\bf 8})$ coupling by the ideally mixed nonet scheme 
\cite{Okubo} or simply the naive quark model. Since the processes are two-body 
decays on mass shell, the amplitude ratios are actually coupling ratios. 
Therefore $\xi$, $\eta$, $\xi'$, and $\eta'$ can be chosen to be real numbers. 
We have only sign ambiguity instead of continuous phase ambiguity, 
when different terms interfere in squared matrix elements.

  A few parameter-independent relations can be read off from Table 1.
\begin{equation}
    |M(b_1^+\pi^-)|^2 = |M(b_1^0\pi^0)|^2.   \label{isospinrule}
\end{equation}
This relation is actually a consequence of charge conjugation invariance and
the isospin property of electromagnetic current alone, not of SU(3) symmetry:
Since $b_1\pi$ is G odd, only the isoscalar part of electromagnetic current
is capable of producing $b_1\pi$ in one-photon annihilation.
Then the relation (\ref{isospinrule}) follows immediately.
Only photon-loop corrections on the light quarks can violate it.
This is one of the special cases where even the photon interaction cannot
violate isospin nor G parity invariance.
Despite the very robust nature of the relation, the current 
data are only marginally consistent with it:
Br$(J/\psi\rightarrow b_1^0\pi^0) = (2.3\pm 0.6)\times 10^{-3}$ {\it vs} 
Br$(J/\psi\rightarrow b_1^+\pi^- + cc)= (3.0\pm 0.5)\times 10^{-3}$\cite{PDG}.
It has been known that the similar equality
Br$(\rho^+\rightarrow \pi^+\gamma)=$ Br$(\rho^0\rightarrow \pi^0\gamma)$ is not 
well satisfied \cite{PDG}. The $\rho-\omega$ mixing is probably responsible
for the violation.  The other parameter-independent relation from Table 1 is:
\begin{equation}
    |M(K_1^+(1270)K^-)|^2 + |M(K_1^+(1400)K^-)|^2 = 
                          |M(b_1^+\pi^-)|^2 + |M(a_1^+\pi^-)|^2.
       \label{ratesumrule}
\end{equation}
To test this sum rule, we need measurement of the $a_1^{\pm}\pi^{\mp}$ mode. 
More interesting is the relation,
\begin{equation}
   |{\rm M}(K_1^0(1400)\overline{K}^0)|^2
   = \tan^2\theta\times |{\rm M}(K_1^0(1270)\overline{K}^0)|^2. \label{K10}
\end{equation}
It will be able to determine $\theta$ directly without referring to other 
parameters. In this paper we focus on
$M(K_1^+(1270)K^-)$ and $M(K_1(1400)^+K^-)$ for $J/\psi$ and 
$\psi'$ on which BES Collaboration shed a light.

\section{Ranges of parameters}

  The decay pattern of the $K_1$ mesons first alerted theorists of the
$K_A-K_B$ mixing.   Earlier theoretical works \cite{old} pointed 
to the maximal mixing of $\theta = 45^{\circ}$.  The maximal mixing 
occurs if the diagonal elements of the $K_A-K_B$ mass matrix are 
exactly equal. Phenomenologically, however, 
the latest decay data still allow for sizable uncertainty \cite{suzuki}:
\begin{equation}
        30^{\circ} < \theta < 60^{\circ}.   \label{thetarange}
\end{equation} 
Since the masses of $K_A$ and $K_B$ are not directly measurable, they
must be computed by theory. Therefore determination of the $K_A-K_B$ diagonal 
mass difference is subject to uncertainties of theoretical assumptions, 
some kinematical and others dynamical. In our analysis below, 
we treat the entire range of Eq.(\ref{thetarange}) as allowed.

  We can make a crude estimate of magnitude of ($\xi, \eta$) and 
($\xi', \eta'$) by comparing the integrated decay rates of 
three-gluon and one-photon annihilation. From measurement \cite{PDG} we know
\begin{eqnarray}
 \frac{\Gamma(1^-(\overline{c}c)\rightarrow\gamma\rightarrow {\rm hadrons})}
{\Gamma(1^-(\overline{c}c)\rightarrow {\rm ggg}\rightarrow {\rm hadrons})}
           & = & 0.25\pm 0.03 \;\;\;\;{\rm for}\; J/\psi, \nonumber \\
           & = & 0.26\pm 0.04 \;\;\;\;{\rm for}\; \psi'. \label{ratio}
\end{eqnarray}
The square roots of the right-hand sides give us an indication of the
amplitude ratios.  There is no compelling reason to equate 
these numbers to $\xi$ and $\eta$, or $\xi'$ and $\eta'$ 
of exclusive channels.  With no better clue at hand, however, 
we use Eq.(\ref{ratio}) to set the ballpark ranges in which the parameter 
values are found. Our parameters are so normalized in Table 1 that the ratio 
of sum of the one-photon rates over all $1^+_B({\bf 8})\,0^-({\bf 8})$ channels 
to sum of the three-gluon rates is equal to $|\eta|^2$.  The normalization of
$\xi$ is chosen in parallel to that of $\eta$. If we equate $|\eta|^2$ to the 
number in the first line of Eq.(\ref{ratio}), we obtain $|\eta|\approx 0.5$. 
It is not unreasonable to expect that $|\xi|$ is in a range similar to that of 
$|\eta|$.  Therefore our very crude estimate or guess is: 
\begin{equation}
   |\xi|, \;\;\; |\eta| \leq  0.5 \;\;\;\;{\rm for}\; J/\psi. 
                 \label{xieta}
\end{equation}
By the same assumption we obtain for $\psi'$
\begin{equation}
   |\xi'|, \;\;\; |\eta'| \leq  0.5  \;\;\;\;{\rm for}\; \psi'.
                  \label{xieta'}
\end{equation}  
In terms of amplitudes, the one-photon process is by no means a small
correction.  In fact, it is known that in some exclusive decay channels 
G parity and/or SU(3) violating amplitudes are comparable to corresponding 
conserved ones.  For instance, we find in the Review of Particle Physics
\cite{PDG} the wrong-to-right G parity amplitude ratio 
$[{\rm Br}(J/\psi\rightarrow\rho\eta')/{\rm Br}(J/\psi\rightarrow\omega\eta')]
^{1/2}\approx 0.8$. We shall keep Eqs.(\ref{xieta}) and, (\ref{xieta'}) 
in mind in the following analysis.   

\section{Analysis of data}

For $J/\psi$ the average of two measurements on the decay 
$J/\psi\rightarrow b_1^{\pm}\pi^{\mp}$ is \cite{PDG}:
\begin{equation}
  {\rm Br}(J/\psi\rightarrow b_1^+\pi^- + cc) = (3.0\pm 0.5)\times 10^{-3},
                       \label{psib}
\end{equation}
The new BES measurements are:
\begin{equation}
   {\rm Br}(J/\psi\rightarrow K_1^+(1270)K^- + cc) < 1.8 \times 10^{-3}
            \;\;\;\; 90\% {\rm CL},
\end{equation}
\begin{equation}
  {\rm Br}(J/\psi\rightarrow K_1^+(1400)K^- + cc) = (5.0\pm 1.3)\times 10^{-3}.
                       \label{psiK1}
\end{equation}
In addition, the BES Collaboration measured three $1^+0^-$ decay modes 
of $\psi'$:
\begin{equation}
   {\rm Br}(\psi'\rightarrow b_1^+\pi^- + cc)= (7.3\pm 1.9)\times 10^{-4},
\end{equation}
\begin{equation}
   {\rm Br}(\psi'\rightarrow K_1^+(1270)K^- + cc) =(7.6\pm 1.7)\times 10^{-4},
\end{equation}   
\begin{equation}
   {\rm Br}(\psi'\rightarrow K_1^+(1400)K^- + cc) < 2.7\times 10^{-4}
             \;\;\;\; 90\% {\rm CL}.
\end{equation}

  The most conspicuous is the pattern that $K_1(1270)^+K^-$ is suppressed
in the $J/\psi$ decay while $K_1(1400)^+K^-$ is suppressed in the $\psi'$
decay.  This is incompatible with the zeroth order picture of the 
maximally mixed $K_A-K_B$ combined with three-gluon annihilation 
dominance. We ask whether inclusion of one-photon annihilation and possibly
a deviation of $\theta$ from $45^{\circ}$ can explain this pattern or not.

\subsection{$J/\psi$ decay}
The ratio of the two $K_1^+K^-$ amplitudes  
can be expressed in our parametrization as
\begin{equation}
   \frac{{\rm M}(K_1^+(1270)K^-)}{{\rm M}(K_1^+(1400)K^-)}
     =\frac{\sqrt{2}\xi\tan\theta+(1+\sqrt{2/5}\eta)}
      {\sqrt{2}\xi-(1+\sqrt{2/5}\eta)\tan\theta}. \label{psiKK}
\end{equation}
The amplitude ratio of $K_1^+(1400)K^-$ to $b_1^+\pi^-$ is subject to the
experimental constraint from Eqs.(\ref{psib}) and (\ref{psiK1}):
\begin{eqnarray}
   \frac{{\rm M}(K_1^+(1400)K^-)}{{\rm M}(b_1^+\pi^-)} & = &
   \frac{\sqrt{2}\xi\cos\theta-(1+\sqrt{2/5}\eta)\sin\theta}
       {1+\sqrt{2/5}\eta},      \\  \label{btoK}
          & = & -1.36 \pm 0.47.     \label{btoKexp}
\end{eqnarray}
The number in the last line has been extracted with the s-wave decay 
assumption. Mixture of d-wave tends to raise the magnitude of the central 
value, for instance, from $-1.36$ to $-1.67$ for $50\%$ mixture of d-wave.  
In order to suppress Br$(J/\psi\rightarrow K_1^+(1270)K^-)$ relative to 
Br$(J/\psi\rightarrow K_1^+(1400)K^-)$, the three-gluon and one-photon 
terms must interfere destructively in the former and constructively 
in the latter. Therefore $\xi$ must be negative according to Eq.(\ref{psiKK}).
Since $\xi$ and $\eta$ enter Eqs.(\ref{psiKK}) and (\ref{btoK}) only through 
the ratio $\xi/(1+\sqrt{2/5}\eta)$, we can eliminate this ratio and
express Eq.(\ref{psiKK}) in terms of $\theta$ and the experimental value 
of Eq.(\ref{btoKexp}). In FIG.1 we have plotted the ratio 
${\rm Br}(J/\psi\rightarrow K_1^+(1270)K^- + cc)/
{\rm Br}(J/\psi\rightarrow K_1^+(1400)K^- + cc)$ as a function of $\tan\theta$.

\begin{figure}
\begin{center}
\mbox{\epsfig{file=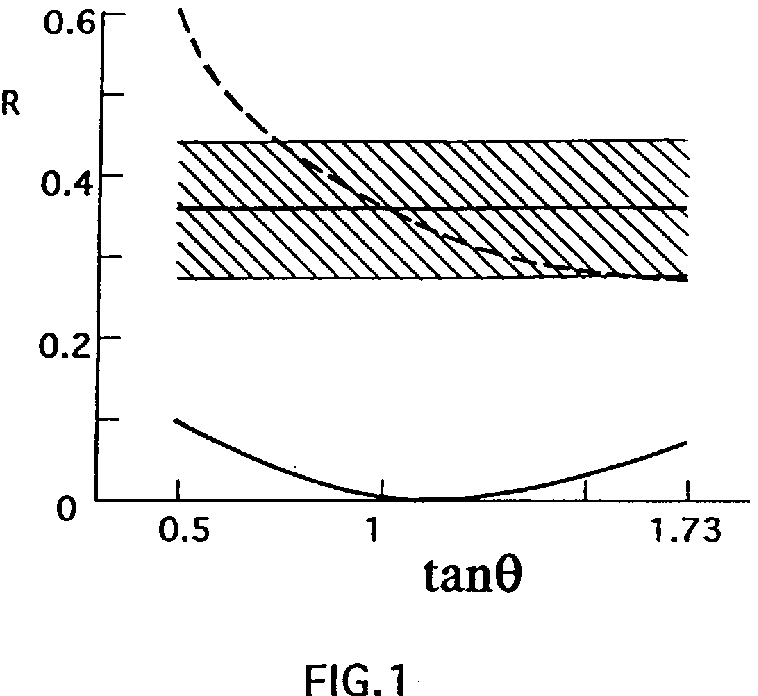,width=7cm,height=5cm}}
\end{center}
\end{figure}

We see that any value of $\theta$ between $30^{\circ}$ and $60^{\circ}$ 
produces a number much smaller than the experimental upper bound,
\begin{equation}
  R\equiv \frac{{\rm Br}(J/\psi\rightarrow K_1^+(1270)K^- + cc)}
        {{\rm Br}(J/\psi\rightarrow K_1^+(1400)K^- + cc)} < 0.36\pm 0.09. 
\end{equation}
We must make sure that we can find values for $\xi$ and $\eta$ 
in the acceptable range of Eq.(\ref{xieta}) under the constraint of 
Eq.(\ref{btoK}). It happens that this constraint
is insensitive to $\theta$ ($=30^{\circ} \sim 60^{\circ}$).  
For illustration we have shown in FIG.2 the range of ($\xi, \eta$)
that correctly produces Eq.(\ref{btoKexp}) for $\theta = 45^{\circ}$.

\begin{figure}
\begin{center}
\mbox{\epsfig{file=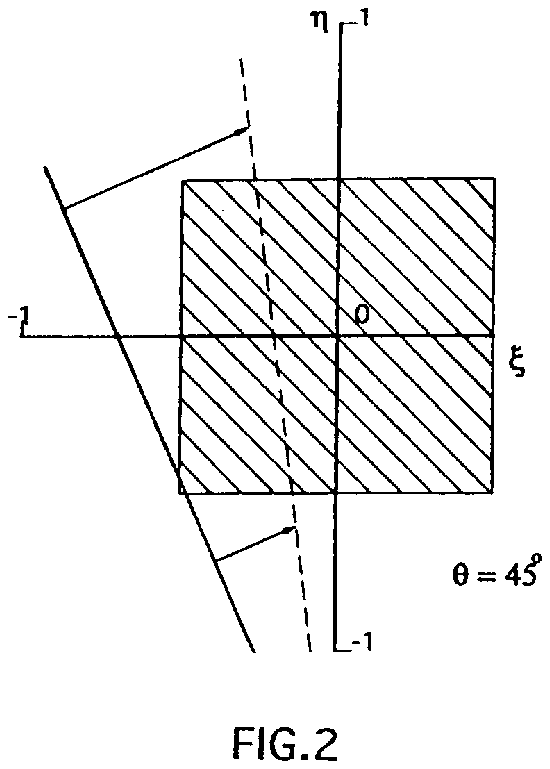,width=5cm,height=7cm}}
\end{center}
\end{figure}
 
Irrespective of values of $\theta$, preferred values for $\xi$ and $\eta$ 
are always near one of the boundary corners ($\xi\approx -0.5$, 
$\eta\approx -0.5$) of Eq.(\ref{xieta}). If we allow for an upward experimental 
error (making smaller in absolute value) in Eq.(\ref{btoKexp}), $\xi$ 
and $\eta$ can be made smaller in magnitude. An upward shift of one 
standard deviation of Eq.(\ref{btoKexp}) would raise the prediction in
Fig.1 to the thin broken curve. 
 
Therefore the BES measurement of the decay $J/\psi\rightarrow K_1^+K^-$ 
can be accommodated with theory once a right amount of the one-photon 
annihilation contribution is added.

\subsection{$\psi'$ decay}
  
We want to be consistent with the measurement
\begin{equation}
 R'\equiv \frac{{\rm Br}(\psi'\rightarrow K_1^+(1400)K^- + cc)}
         {{\rm Br}(\psi'\rightarrow K_1^+(1270)K^- + cc)} < 0.36\pm 0.08,
             \label{psi'KKrate}
\end{equation}
with the matrix element ratio,
\begin{equation}
  \frac{{\rm M}(K_1^+(1400)K^-)}{{\rm M}(K_1^+(1270)K^-)} =
  \frac{\sqrt{2}\xi'-(1+\sqrt{2/5}\eta')\tan\theta}
       {\sqrt{2}\xi'\tan\theta+(1+\sqrt{2/5}\eta')}.
              \label{psi'KK} 
\end{equation} 
Positive $\xi'$ suppresses the numerator by destructive interference.    
We have the experimental information
on the ratio of $K_1^+(1270)K^-$ to $b_1^+\pi^-$:
\begin{eqnarray}
 \frac{{\rm M}(\psi'\rightarrow K_1^+(1270)K^-)}
      {{\rm M}(\psi'\rightarrow b_1^+\pi^-)} & = &
 \frac{\sqrt{2}\xi'\sin\theta+(1+\sqrt{2/5}\eta')\cos\theta}
      {1+\sqrt{2/5}\eta'}    \\
       & = & 1.03 \pm 0.34.         \label{btoK'}
\end{eqnarray}
Again expressing the ratio Eq.(\ref{psi'KK}) as a function of $\theta$ alone, 
we have plotted the ratio $R'$ in FIG.3.

\begin{figure}
\begin{center}
\mbox{\epsfig{file=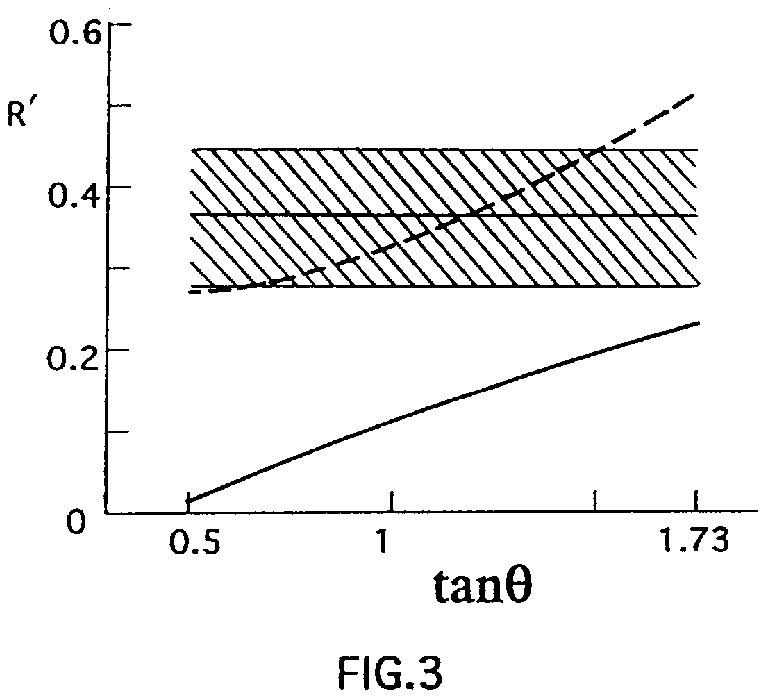,width=7cm,height=5cm}}
\end{center}
\end{figure}

 We obtain the solid 
curve for $R'$ when we take the central value in Eq.(\ref{btoK'}).  
As in the case of $J/\psi$, any value of $\theta$ between $30^{\circ}$ 
and $60^{\circ}$ is consistent with the experimental upper bound on $R'$.
Values of $\xi'$ and $\eta'$ are constrained by
the ratio of $K_1^+(1270)K^-$ to $b_1^+\pi^-$ of Eq.(\ref{btoK'}).

\begin{figure}
\begin{center}
\mbox{\epsfig{file=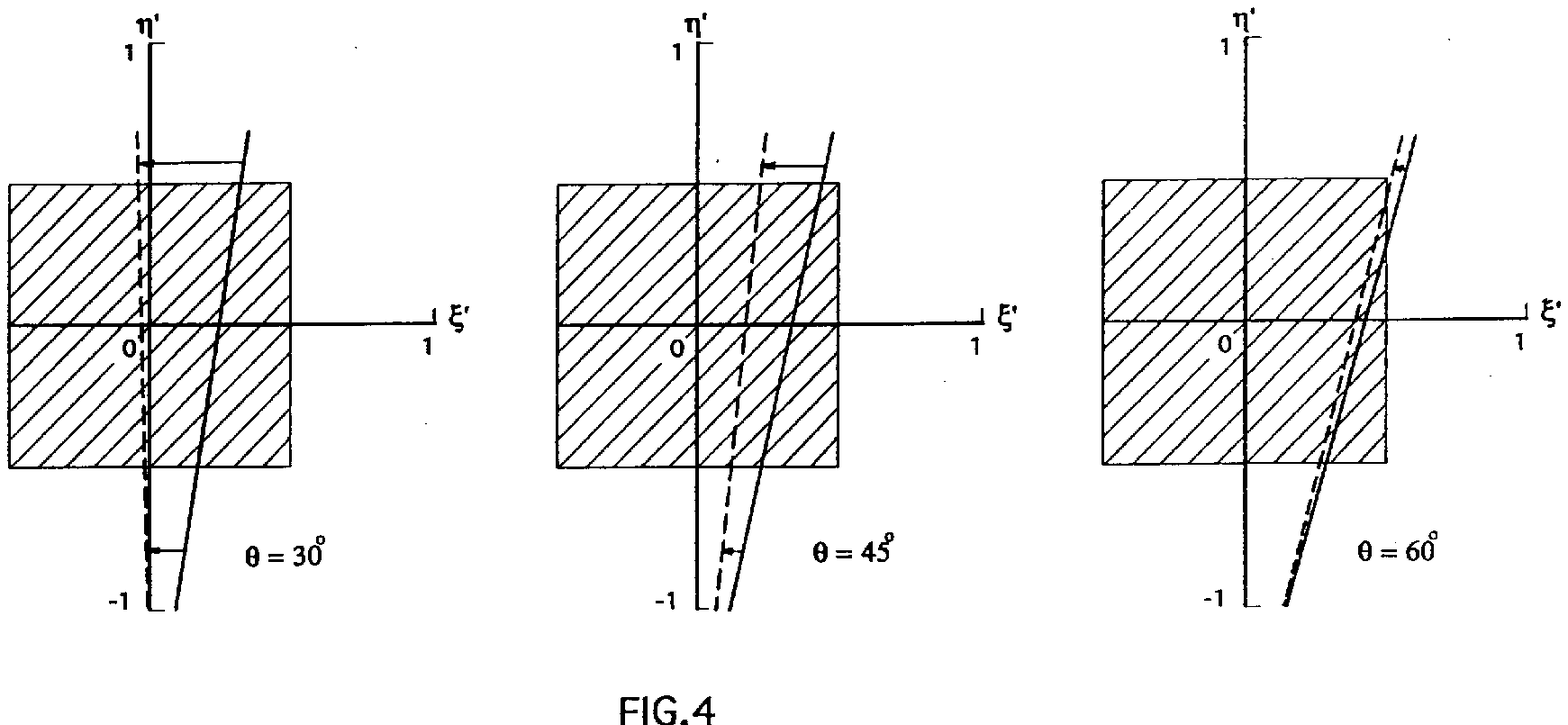,width=12cm,height=7cm}}
\end{center}
\end{figure}

FIG.4 depicts the allowed ranges for $\xi'$ and $\eta'$ when 
$\theta=30^{\circ}, 45^{\circ}$, and $60^{\circ}$. 
$\xi'$ and $\eta'$ must be on the solid straight line 
when the central value is taken in Eq.(\ref{btoK'}).  This time the
constraint on $\xi'$ and $\eta'$ are mildly dependent on $\theta$.
Though we can find $\xi'$ and $\eta'$ in the acceptable range (\ref{xieta'})
for any value of $\theta$, $\xi'$ can be smaller when 
$\theta < 45^{\circ}$ than when $\theta > 45^{\circ}$.
If we make a downward shift of the number in Eq.(\ref{btoK'}), even smaller  
values would be allowed for $\xi'$. However we can make such a shift only by 
a half standard deviation or less, depending on $\theta$, in order to be
compatible with the upper bound on $R'$.  The situation is shown in FIG.3 and
FIG.4.

 Let us summarize our numerical analysis: After a reasonable amount of the
one-photon annihilation amplitudes is included, any value of $\theta$ 
between $30^{\circ}$ and $60^{\circ}$ can be consistent with the $1^+0^-$ 
decay modes of both $J/\psi$ and $\psi'$ that have been so far measured.
No stringent constraint is imposed on the one-photon annihilation amplitudes
of those decay modes either.

Therefore the characteristic of the BES measurement that may look surprising 
at the first sight is not a surprise at all.

\section{Outlook}

   When future experiment measures the neutral modes of $K_1\overline{K}$ and
$\overline{K}_1 K$, they will allow us to restrict the ranges of parameters 
more stringently. Our SU(3) parametrization gives us
\begin{eqnarray}
 \frac{{\rm Br}(J/\psi\rightarrow K_1^0(1400)\overline{K}^0 + cc)}
      {{\rm Br}(J/\psi\rightarrow K_1^+(1400)K^- + cc) } 
      & = & \biggl|\frac{(1-\sqrt{8/5}\eta)\tan\theta}
            {\sqrt{2}\xi-(1+\sqrt{2/5}\eta)\tan\theta}\biggr|^2,\\
                    \label{psiprediction} 
 \frac{{\rm Br}(\psi'\rightarrow K_1^0(1270)\overline{K}^0 + cc)}
      {{\rm Br}(\psi'\rightarrow K_1^+(1270)K^- + cc)} 
      & = & \biggl|\frac{1-\sqrt{8/5}\eta'}
                 {\sqrt{2}\xi'\tan\theta+(1+\sqrt{2/5}\eta')}\biggr|^2.
                    \label{psi'prediction}
\end{eqnarray}
As parameters sweep in the currently allowed region, these ratios 
change over a wide range both above and below unity.   For the 
purpose of fixing the mixing angle, the ratio 
${\rm Br}(K_1^0(1400)\overline{K}^0)/{\rm Br}(K_1^0(1270)\overline{K}^0)$ 
is by far the cleanest source (cf. Eq.(\ref{K10})). 
As for the $b_1\pi$ modes, ${\rm Br}(b_1^0\pi^0)={\rm Br}(b_1^+\pi^-)$ is 
a robust prediction which follows from the isospin and charge conjugation
property of the electromagnetic current. 
Violation of this equality would mean that emission and 
reabsorption of a photon somewhere inside the light quark sector is enhanced.
If $\rho-\omega$ mixing should cause a substantial misidentification of
$a_1$ and $b_1$, we would have to modify our analysis
by including this $b_1-a_1$ mixing. 

To draw a definite conclusion on the $K_A-K_B$ mixing from the
$1^+0^-$ decays of orthocharmonia, we shall have to wait until
the neutral modes $J/\psi\rightarrow K_1^0\overline{K}^0+cc$, 
$\psi'\rightarrow K_1^0\overline{K}^0+cc$ \cite{Olsen}
and $\psi'\rightarrow b_1^0\pi^0$ are measured. 
In addition, a more accurate measurement
of the $b_1^0\pi^0/b_1^+\pi^-$ ratio in the $J/\psi$ decay is desired.

\vskip 1cm
\noindent
{\large \bf Acknowledgements}
\vskip 0.5cm
I thank S. Olsen for calling my attention to the new measurement of the
BES Collaboration and for the subsequent communications that prompted me 
to go through this analysis.
This work was supported in part by National Science
Foundation under grant PHY--95--14797 and
in part by the Director, Office of Energy Research, Office of
High Energy and Nuclear Physics, Division of High Energy Physics of the
U.S. Department of Energy under Contract DE--AC03--76SF00098.


\newpage

\begin{table}
\caption{SU(3) parametrization of decay amplitudes for
$J/\psi\rightarrow 1^+ 0^-$.  
The decay modes not listed below are forbidden by charge 
conjugation invariance or the Okubo-Zweig-Iizuka rule.
For the $\psi'$ decay, replace $\xi$ and $\eta$ by
$\xi'$ and $\eta'$, respectively.}

\vskip 1cm

\begin{tabular}{|c|c|}  
     $a_1^+\pi^- (= - a_1^-\pi^+)$    & $\sqrt{2}\xi$   \\
     $K_A^+K^- (= -K_A^-K^+)$         & $\sqrt{2}\xi$   \\
     $K_A^0\overline{K}^0 (=-\overline{K}_A^0K^0)$  & 0  \\  \hline
     $b_1^+\pi^- (=b_1^-\pi^+)$       & $1 + \sqrt{2/5}\eta$ \\
     $b_1^0\pi^0$                     & $1 + \sqrt{2/5}\eta$ \\
     $b_1^0\eta$                      & $\;\;\;\;\sqrt{6/5}\eta$  \\  
     $K_B^+K^- (=K_B^-K^+)$           & $1 + \sqrt{2/5}\eta$  \\
     $K_B^0\overline{K}^0 (=\overline{K}_B^0K^0$)  & $1-\sqrt{8/5}\eta$ \\
     $h_1\eta$                        & $\;\;\;\;\sqrt{1/3}+\sqrt{2/15}\eta$\\
     $h_1\pi^0$                       & $\sqrt{18/5}\eta$ \\
     $h'_1\eta$                       & $\sqrt{2/3}-(4/\sqrt{15})\eta$  
\end{tabular}
\label{table:1}
\end{table}

\vskip 2cm

\noindent
\begin{figure}
\caption{$R={\rm Br}(J/\psi\rightarrow K_1^+(1270)K^-+cc)/
{\rm Br}(J/\psi\rightarrow K_1^+(1400)K^-+cc)$ against $\tan\theta$.
The shaded band is the experimental upper bound when one standard deviation
error is taken into account. The broken curve is for 
${\rm M}(K_1^+(1400)K^-)/{\rm M}(b_1^+\pi^-)
= -1.36 + 0.47$.\label{fig:1}}
\vskip .5in

\caption{The range of $\xi$ and $\eta$ allowed by Eq.(23) for 
$\theta =45^{\circ}$. The shaded region is $|\xi|, |\eta| < 0.5$. 
$\xi$ and $\eta$ are constrained on the solid line when the central value is 
taken in Eq.(23). This solid line is virtually independent of $\theta$ between
$30^{\circ}$ and $60^{\circ}$. The broken line is for ${\rm M}(K_1^+(1400)K^-)
/{\rm M}(b_1^+\pi^-)= -1.36+0.47$. \label{fig:2}}
\vskip .5in

\caption{$R'={\rm Br}(\psi'\rightarrow K_1^+(1400)K^-+cc)/
{\rm Br}(\psi'\rightarrow K_1^+(1270)K^-+cc)$ against $\tan\theta$.
The shaded band is the experimental upper bound with one standard deviation
error. The solid curve below the band is the prediction when the central value
is taken in Eq.(28). It rises to the broken curve passing through the band when
the central value is lowered by a half standard deviation to $1.03 -0.17$
in Eq.(28). \label{fig:3}}
\vskip .5in

\caption{The ranges of $\xi'$ and $\eta'$ for $\theta= 30^{\circ}, 45^{\circ}$,
and $60^{\circ}$.  The shaded region is $|\xi'|, |\eta'| < 0.5$.
$\xi'$ and $\eta'$ are constrained on the solid line when the
central value is taken in Eq.(28).  
We can reduce magnitude of $\xi'$ and $\eta'$ 
without conflicting $R'$ by lowering the number in Eq.(28), but no more than 
a half standard deviation (cf. FIG. 3). The broken curve is the limit to
which we can move the constraint without conflicting with $R'$.\label{fig:4}}
\end{figure}


\begin{thebibliography}{99}
\bibitem{old}   E. W. Colglazier and J. L. Rosner, Nucl. Phys. {\bf B27}, 349
        (1971): G. W. Brandenburg {\it et al}., Phys. Rev. Lett. {\bf 36}, 703
        (1976): R. K. Carnegie {\it et al}., Nucl. Phys. {\bf B127},
        509 (1977): R. K. Carnegie {\it et al}., Phys. Lett. {\bf 68B}, 
        289 (1977): H. J. Lipkin, Phys. Lett. {\bf 72B}, 249 (1977).
\bibitem{suzuki} M. Suzuki, Phys. Rev. D {\bf 47}, 1252 (1993). 
\bibitem{new} H. G. Blundel, S. Godfrey, and B. Phelps, Phys. Rev. D {\bf 53},
          3712 (1996).
\bibitem{BES} BES Collaboration, Report to the XXVIII International 
                 Conference on High Energy Physics, July 25-31, 1996, 
                 Warsaw, Poland, to be published.
\bibitem{PDG} Particle Data Group, R. M. Barnett et al, 
              Phys. Rev. D {\bf 54}, 1 (1996).
\bibitem{Okubo}  S. Okubo, Phys. Lett. {\bf 5}, 165 (1963).
\bibitem{Olsen}  The BES Collaboration is currently working toward 
              determination of the branching fractions for the 
              $K_1^0\overline{K}^0 + \overline{K}_1^0K^0$ channels. 
              S. Olsen, a private communication.


\end{thebibliography}
\end{document}